\documentclass[prl,twocolumn,amsmath,amssymb,floatfix]{revtex4}
\usepackage{graphicx}

\usepackage{epsfig}
\usepackage{dcolumn}
\usepackage{bm}

\def\(({\left(}
\def\)){\right)}                       
\def\[[{\left[}
\def\]]{\right]}    
\newcommand{\be}{\begin{equation}}
\newcommand{\ee}{\end{equation}}
\newcommand{\bea}{\begin{eqnarray}}
\newcommand{\eea}{\end{eqnarray}}

\begin{document} 
\title{Comment on ``Large Slip of Aqueous Liquid Flow over a Nanoengineered Superhydrophobic
Surface''}
\author{Lyd\'eric Bocquet$^{1}$, Patrick Tabeling$^{2}$, S\'ebastien Manneville$^{3}$}
\affiliation{(1) LPMCN, Universit\'e Lyon I, UMR CNRS 5586, Lyon, France\\
(2) PCT, ESPCI, UMR CNRS 7083, Paris, France\\
(3) CRPP, UPR CNRS 8641, Pessac, France}



\maketitle
In a recent Letter \cite{CK06}, Choi and Kim reported slip lengths of a few tens of
microns for water on nanoengineered superhydrophobic surfaces, on the basis of
rheometry (cone-and-plate) measurements.
We show that the experimental uncertainty
in the experiment of Ref. \cite{CK06}, expressed in term of slip lengths, lies in the range
20 - 100 micrometers, which is precisely the order of magnitude of the
reported slip lengths. 
Moreover we point out a systematic bias expected on the
superhydrophobic surfaces. We thus infer that it is not possible to draw out
any conclusion concerning the existence of huge slip lengths in the system
studied  by Choi and Kim.

Choi and Kim performed torque measurements using a commercial rheometer (AR2000, TA Instruments) with a cone-and-plate geometry. The slip length $\delta$ is deduced from the correction of the torque $M$ w.r.t. a reference no-slip value $M_0$~:
${\delta}={2\over 3}\theta_0 R (1-M/M_0)$.
$M_0={2\pi\over 3} \mu \Omega R^3/\theta_0$ is the prediction in the
absence of slippage; $R$ is the cone radius ($\simeq 3$ cm) and $\theta_0$ its opening angle ($2^\circ$);
 $\mu$ is the viscosity of the liquid; $\Omega$ is the rotational velocity. 
Putting numbers in this expression  shows that the reported $20\mu$m slip length for water
corresponds to a  3\% correction to the (small) reference  torque  
$M_0 \sim 5\mu$Nm. 
The authors however claim a $3 \mu $m 
uncertainty on the slip length, which  corresponds to an overall $0.5$\% errorbar on the {\it relative
deviation} of the torque $(M_0-M)/M_0$.
Such a precision is not attainable in the present experiment. To illustrate this uncertainty issue, we have performed benchmark experiments using an AR2000 rheometer with a smooth, stainless steel cone-and-plate geometry 
(with the same radius and cone angle), very close to that of Ref.~\cite{CK06}. 
This rheometer was calibrated using a reference Newtonian silicon oil 
(BR0050CPS, $\mu=48.4$~mPa\,s at 25$^\circ$C), which yielded the expected value to 
within 0.5\%. Then, turning to 
distilled deionized water at 25$^\circ$C, we performed torque measurements similar to  \cite{CK06}
and measured the torque standard deviation $\Delta M/M$ with this liquid,
yielding 
$\Delta M/M\simeq$~1.4\%
at $\dot\gamma=150$ s$^{-1}$, up to 3.5\% at $\dot\gamma=50$~s$^{-1}$. This is 
far above the claimed $0.5$\% uncertainty.
If the uncertainties on the filling volume and on the gap size are 
included, the global uncertainty on the measured torque is at least $\Delta M/M\simeq$~3\%. 
Finally, using the expression $\delta(M)$ and adding a similar uncertainty on the reference viscosity $\mu$ in $M_0$, 
leads to $\Delta\delta/\delta\simeq$~100-200\%, 
so that $\Delta\delta \sim 20-40 \mu$m for water and $\Delta\delta \sim 50-100 \mu$m for glycerin.
The reported effect are therefore {\it within uncertainty} and the experiment of Ref. \cite{CK06} should be considered as inconclusive.

Another source of
difficulties in the interpretation of the experiment in Ref \cite{CK06} is the role of secondary
flows. 
The relevant reynolds number is ${\cal R}e= \rho \Omega R^2 \theta_0^2/\mu  \gtrsim 2$, 
and inertial effects should lead to an increase in the torque up to two percent ($ M-M_0= 6. 10^{-4} {\cal R}e^2$
 \cite{Walter}). In view of the claimed resolution, this correction should be measurable in the
experiment of Ref. \cite{CK06} and interpreted as an apparent negative slip length 
up to $\sim -15 \mu$m for the smooth hydrophilic surfaces. This effect is not detected 
in \cite{CK06}, which furthermore confirms 
the weakness in the interpretation of the measurements. 
At this stage, it is worth pointing out a {\it systematic bias} 
on the superhydrophobic surfaces.
Indeed, for the same liquid volume
filling the gap in the cone-and-plate, 
the meniscus at the edges makes the radius $R$ slightly
smaller on the super-hydrophobic surface
(with very large contact angle), w.r.t. the other surfaces with smaller contact angles.
The variation in $R$ is predicted to be 
of the order of a fraction of the gap at the border $\Delta R\approx - \alpha \theta_0 R$ (with $\alpha \lesssim1$).
Assuming no-slip at the surfaces, the resulting decrease of the solid-liquid area leads to a reduction of torque on the superhydrophobic surfaces,
$\Delta M_{cap}/M_0\approx - 3\alpha \theta_0$, of the order of a few percents.
The misinterpretation of this effect using the equation for $\delta(M)$ thus erroneously predicts a slippage effect with a slip length $\delta_{\rm eff}\approx 2 \alpha \theta_0^2 R$, of the order of a few tens of micrometers. See also note \cite{note}.

In summary, the experimental
uncertainty that we estimate is comparable to the amplitude of the effect the
authors have observed. Moreover
a systematic bias could be wrongly interpreted in terms of very large slippage on 
superhydrophobic surfaces. The experiments of Ref. \cite{CK06}Ê
are therefore inconclusive. 


\begin{references}
\bibitem{note} {To fix the ideas, a retreat of $\sim 0.2$ mm in the average position of the
edge meniscus could be misinterpreted as a slip length increase of $\sim 14$
microns. This underlines the importance of an effect which was erronously
overlooked in Ref \cite{CK06}.}
\end{references}
\end{document}